\documentclass[prd,aps,showpacs,nofootinbib,eqsecnum,onecolumn,groupedaddress,amssymb]{revtex4}
\usepackage{amsmath}
\usepackage{amssymb}
\usepackage{amsfonts}
\usepackage{graphicx,bm}
\usepackage{dcolumn}
\usepackage{color,amsxtra}
\usepackage{epsf}
\usepackage{enumerate}
\usepackage{hhline}
\usepackage{array}
\usepackage{tabularx}
\usepackage{subfigure}
\usepackage{fancyhdr}
\usepackage{mathrsfs}

\newcommand{\be}{\begin{equation}}
\newcommand{\ee}{\end{equation}}
\newcommand{\bea}{\begin{eqnarray}}
\newcommand{\eea}{\end{eqnarray}}
\newcommand{\beaa}{\begin{eqnarray*}}
\newcommand{\eeaa}{\end{eqnarray*}}

\newcommand{\nn}{\nonumber \\}

\allowdisplaybreaks[4]

\newcommand{\Eqn}[1]{&\hspace{-0.2em}#1\hspace{-0.2em}&}

\def\be{\begin{equation}}
\def\ee{\end{equation}}
\def\bea{\begin{eqnarray}}
\def\eea{\end{eqnarray}}

\def\nn{\nonumber \\}

\begin{document}

\title{Reconstruction of scalar field theories realizing inflation consistent with the Planck and BICEP2 results}

\author{Kazuharu Bamba$^{1, 2}$, Shin'ichi Nojiri$^{3, 4}$
and Sergei D. Odintsov$^{5, 6, 7}$
}
\affiliation{
$^1$Leading Graduate School Promotion Center,
Ochanomizu University, 2-1-1 Ohtsuka, Bunkyo-ku, Tokyo 112-8610, Japan\\
$^2$Department of Physics, Graduate School of Humanities and Sciences, Ochanomizu University, Tokyo 112-8610, Japan\\
$^3$Kobayashi-Maskawa Institute for the Origin of Particles and the
Universe, Nagoya University, Nagoya 464-8602, Japan\\
$^4$Department of Physics, Nagoya University, Nagoya 464-8602, Japan\\
$^5$Consejo Superior de Investigaciones Cient\'{\i}ficas, ICE/CSIC-IEEC, 
Campus UAB, Facultat de Ci\`{e}ncies, Torre C5-Parell-2a pl, E-08193
Bellaterra (Barcelona), Spain\\
$^6$Instituci\'{o} Catalana de Recerca i Estudis Avan\c{c}ats
(ICREA), Barcelona, Spain\\ 
$^7$Tomsk State Pedagogical University, 634061 Tomsk and 
National Research Tomsk State University, 634050 Tomsk, Russia
}


\begin{abstract}
We reconstruct scalar field theories to realize 
inflation compatible with the BICEP2 result as well as the Planck. 
In particular, we examine the chaotic inflation model, natural (or axion) inflation model, and an inflationary model with a hyperbolic inflaton potential. 
We perform an explicit approach to find out a scalar field model of inflation in which any observations can be explained in principle. 
\end{abstract}

\pacs{98.80.-k, 98.80.Cq, 12.60.-i}
\hspace{14.5cm} OCHA-PP-323

\maketitle

\def\thesection{\Roman{section}}
\def\theequation{\Roman{section}.\arabic{equation}}

\section{Introduction}

In addition to cosmological observations to assist the current accelerated 
expansion of the universe, the so-called dark energy problem (for recent reviews, see~\cite{R-NO-CF-CD, Bamba:2012cp}), 
thanks to the observational data obtained from the recent BICEP2 experiment~\cite{Ade:2014xna}, the Planck satellite~\cite{Ade:2013lta, Ade:2013uln} as well as the Wilkinson Microwave anisotropy probe (WMAP)~\cite{WMAP, Hinshaw:2012aka}, 
the studies on inflation in the early universe have recently been 
executed much more extensively. 

In particular, there have been proposed inflationary models, especially, 
scalar field theories, to fit the observational data obtained from the BICEP2 experiment~\cite{Ade:2014xna} as well as the Planck satellite~\cite{Ade:2013lta, Ade:2013uln} including the running of the 
spectral index for the scalar modes of the primordial 
density perturbations in Refs.~\cite{Hazra:2014goa, Wan:2014fra}\footnote{For an approach to the model building based on the modification of gravity,
 see, e.g.,~\cite{BMOS-BCOZ, Lahanas:2014ula, Takeda:2014qma}.} (for a review on the relation between the form of the inflaton potential and the power spectrum of the density perturbations, see~\cite{Lidsey:1995np}).

In this Letter, we reconstruct the scalar field theories 
to explain the BICEP2 and Planck data. 
Very recently, conditions that a model is consistent with the BICEP2 
result have been examined in Refs.~\cite{Hossain:2014ova, Garcia-Bellido:2014eva, Barranco:2014ira, Martin:2014lra, Gao:2014pca, CH-CHZ, Hamada:2014xka, KS-KSY-HKSY, Hu:2014aua, Sloth:2014sga, Ashoorioon:2014nta}. 
Our aim in this work is to reconstruct wider classes of scalar field theories with inflation. It is also checked whether indeed the running of the spectral index exists and the models reconstructed may be compatible with the BICEP and Planck data. Particularly, we investigate 
the so-called chaotic inflation model~\cite{Linde:1981mu}, natural (or recently called as axion) inflation model~\cite{Freese:1990rb}, 
and an inflationary model where the inflaton has a hyperbolic 
potential, which was first proposed in Ref.~\cite{Yokoyama:1987an}. 
In this way, we can provide the method to build the scalar field 
models fitting any data of future observations different from the BICEP and Planck. 
We use units of $k_\mathrm{B} = c = \hbar = 1$ and express the
gravitational constant $8 \pi G$ by
${\kappa}^2 \equiv 8\pi/{M_{\mathrm{Pl}}}^2$ 
with the Planck mass of $M_{\mathrm{Pl}} = G^{-1/2} = 1.2 \times 
10^{19}$\,\,GeV. 

The Letter is organized as follows. 
In Sec.~II, we describe the reconstruction of the scalar field theory. 
In Sec.~III, we present the scalar field models compatible with the Planck and 
BICEP2 data. Summary is finally presented in Sec.~IV.

\section{Reconstruction of scalar field theories}

We study the model of a scalar field coupled with gravity 
\be
\label{S1}
S = \int d^4 x \sqrt{-g} \left( \frac{R}{2\kappa^2} 
 - \frac{1}{2}\partial_\mu \phi \partial^\mu \phi - V(\phi) \right)\, .
\ee
The slow-roll parameters $\epsilon$, $\eta$ and $\xi$ are given by 
\be
\label{S2}
\epsilon \equiv \frac{1}{2\kappa^2} \left( \frac{V'(\phi)}{V(\phi)} \right)^2\, ,\quad 
\eta \equiv \frac{1}{\kappa^2} \frac{V''(\phi)}{V(\phi)}\, , \quad 
\xi^2 \equiv \frac{1}{\kappa^4} \frac{V'(\phi) V'''(\phi)}{V(\phi)^2}\, . 
\ee
In the following, the prime denotes the derivative with respect to 
the argument of functions such as $V' (\phi) \equiv \partial V(\phi)/\partial \phi$. 
The tensor-to-scalar ratio is obtained as 
\be
\label{S3}
r = 16 \epsilon\, ,
\ee
%
and the spectral index $n_\mathrm{s}$ of the primordial curvature fluctuations and the associated running of the spectral index $\alpha_\mathrm{s}$ are 
\be
\label{SS1}
n_\mathrm{s} - 1 \sim - 6 \epsilon + 2 \eta\, , \quad 
\alpha_\mathrm{s} \equiv \frac{d n_\mathrm{s}}{d \ln k} 
\sim 16\epsilon \eta - 24 \epsilon^2 - 2 \xi^2
\, .
\ee
%
%
For a detailed review of inflation including the expiations of the above slow-roll parameters, see, for instance,~\cite{Lyth:1998xn}. 
The number of $e$-folds $N$ or the redshift $z$ dependence or the slow-roll parameter $\epsilon$ or the tensor-to-scalar ratio $r$ will be determined by the future observations. Then we now explore how we can construct a model which reproduces the observed values of the slow-roll parameter $\epsilon$ or the tensor-to-scalar ratio $r$. 

We take the flat Friedmann-Lema\^{i}tre-Robertson-Walker (FLRW) metric 
%
$
ds^2 = -dt^2 + a^2(t) \sum_{i=1,2,3}\left(dx^i\right)^2 
$. 
%
Here, $a(t)$ is the scale factor and the Hubble parameter is 
defined as $H \equiv \dot{a}/a$ with the dot showing 
the time derivative. 

The gravitational field equations in the FLRW background are given by 
\be
\label{S4}
\frac{3}{\kappa^2} H^2 = \frac{1}{2}{\dot \phi}^2 + V(\phi)\, , \quad 
 - \frac{1}{\kappa^2} \left( 3 H^2 + 2\dot H \right) 
= \frac{1}{2}{\dot \phi}^2 - V(\phi)\, .
\ee
We may redefine the scalar field $\phi$ by a new scalar field $\varphi$, 
$\phi = \phi (\varphi)$, and identify $\varphi$ as the number of 
$e$-folds $N$. Hence, gravitational field equations (\ref{S4}) 
can be written as 
\be
\label{S5}
\frac{3}{\kappa^2} H(N)^2 = \frac{1}{2}\omega(\varphi) H(N)^2 
+ V\left(\phi\left( \varphi \right) \right)\, , \quad 
 - \frac{1}{\kappa^2} \left( 3 H(N)^2 + 2H'(N) H(N) \right) 
= \frac{1}{2}\omega(\varphi) H(N)^2 
- V\left(\phi\left( \varphi \right) \right)\, , 
\ee
with $\omega(\varphi) \equiv \left( d\phi/d\varphi \right)^2$. 
It is mentioned that the reconstruction of such a class of models 
yielding the equations in (\ref{S5}) was first 
suggested in Ref.~\cite{Nojiri:2005pu}. 
Thus, if the Hubble parameter $H$ is represented as a function of $N$ as 
$H(N)$, and if $\omega(\varphi)$ and $V(\varphi)\equiv V\left(\phi\left(\varphi\right)\right)$ become 
\be
\label{S6}
\omega(\varphi) = - \left. \frac{2 H'(N)}{\kappa^2 H(N)}\right|_{N=\varphi}\, , \quad 
V(\varphi) = \left. \frac{1}{\kappa^2}\left( 3 H(N)^2 + H(N) H'(N) \right) 
\right|_{N=\varphi}\, ,
\ee
we have $H=H(N)$, $\varphi = N$ as a solution for the equation of motion of 
$\phi$ or $\varphi$ and the Einstein equation. 
We should note $H'<0$ because $\omega(\varphi)>0$. 

We here express the slow-roll parameters $\epsilon$, $\eta$ and $\xi$ by $H$ 
as follows 
\begin{align}
\label{S7}
\epsilon =& \left. \frac{1}{2\kappa^2} \left(\frac{d\varphi}{d\phi} \right)^2 
\left( \frac{V'(\varphi)}{V(\varphi)} \right)^2 \right|_{\varphi=N}
= \left. \frac{1}{2\kappa^2} \frac{1}{\omega(\varphi)} 
\left( \frac{V'(\varphi)}{V(\varphi)} \right)^2 \right|_{\varphi=N} \nn
=& - \frac{H(N)}{4 H'(N)} \left[ \frac{6\frac{H'(N)}{H(N)} 
+ \frac{H''(N)}{H(\phi)} + \left( \frac{H'(N)}{H(N)} \right)^2}
{3 + \frac{H'(N)}{H(N)}} \right]^2 \, , \nn
\eta = & \left. \frac{1}{\kappa^2 V(\varphi)} \left[ \frac{d\varphi}{d\phi} 
\frac{d}{d\varphi} 
\left( \frac{d\varphi}{d\phi} \right) V'(\varphi) 
+ \left( \frac{d\varphi}{d\phi} \right)^2 V''(\varphi) \right] \right|_{\varphi=N} 
=  \left. \frac{1}{\kappa^2 V(\varphi)} \left[ - \frac{\omega'(\varphi)}{2 \omega(\varphi)^2} 
V'(\varphi) + \frac{1}{\omega(\varphi)} V''(\varphi) \right] \right|_{\varphi=N} \nn
= & -\frac{1}{2} \left( 3 + \frac{H'(N)}{H(N)} \right)^{-1} \left[ 
9 \frac{H'(N)}{H(N)} + 3 \frac{H''(N)}{H(N)} + \frac{1}{2} \left( \frac{H'(N)}{H(N)} \right)^2 -\frac{1}{2} \left( \frac{H''(N)}{H'(N)} \right)^2 
+ 3 \frac{H''(N)}{H'(N)} + \frac{H'''(N)}{H'(N)} \right] \, , \nn
\xi^2 = & \left. \frac{V'(\varphi)}{\kappa^4 V(\varphi)^2 \omega(\varphi)^2 }
\left\{ \left[ - \frac{\omega''(\varphi)}{2\omega(\varphi)} 
+ \left( \frac{\omega'(\varphi)}{\omega(\varphi)} \right)^2 \right]V' (\varphi) 
 - \frac{3\omega'(\varphi)}{2\omega(\varphi)} V''(\varphi) + V'''(\varphi) 
\right\} \right|_{\varphi=N} \nn
= & \frac{ 6 \frac{H'(N)}{H(N)} + \frac{H''(N)}{H(N)} 
+ \left( \frac{H'(N)}{H(N)} \right)^2 }{4 \left( 3 + \frac{H'(N)}{H(N)} \right)^2} 
\left[ 3 \frac{H(N) H'''(N)}{H'(N)^2} + 9 \frac{H'(N)}{H(N)} 
- 2 \frac{H(N) H''(N) H'''(N)}{H'(N)^3} + 4 \frac{H''(N)}{H(N)} 
\right. 
\nn
& \left. 
{}+ \frac{H(N) H''(N)^3}{H'(N)^4} + 5 \frac{H'''(N)}{H'(N)} 
- 3 \frac{H(N) H''(N)^2}{H'(N)^3} - \left( \frac{H''(N)}{H'(N)} \right)^2 
+ 15 \frac{H''(N)}{H'(N)} 
+ \frac{H(N) H''''(N)}{H'(N)^2} \right]\, .
\end{align}
%
%
%
When we solve equations in (\ref{S7}) with respect to $H(N)$, we can find the corresponding scalar field theory by using (\ref{S6}), in principle. 
It is not so straightforward to solve equations in (\ref{S7}). Therefore, 
we investigate the following case 
\be
\label{S8}
H'''(N) \ll H''(N) \ll H'(N) \ll H(N)\, .
\ee
In this case, we acquire 
\be
\label{S9}
\epsilon(N) \sim - \frac{H'(N)}{H(N)}\quad \mbox{or} \quad 
H(N) \sim H_0 \exp \left( - \int^{N} d \hat{N} \epsilon(\hat{N}) \right)\, ,
\ee
with a constant $H_0$ and also 
\be
\label{SS2}
\eta(N)  \sim - \frac{3}{2} \frac{H'(N)}{H(N)} \sim \frac{3}{2} \epsilon(N)\, , \quad 
\xi^2 \sim \frac{3}{2} \left( \frac{H'(N)}{H(N)} \right)^2 
\sim \frac{3}{2} \epsilon^2 \, . 
\ee
Thus, by using (\ref{S7}), we have 
\be
\label{S10}
\omega (\varphi) \sim  \left. \frac{2}{\kappa^2} \epsilon(N) \right|_{N=\varphi}\, , \quad 
V(\varphi) \sim  \left. \frac{3H_0^2}{\kappa^2} \exp \left( - 2 \int^{N} d \hat{N} \epsilon(\hat{N}) \right)
\right|_{N=\varphi} \, .
\ee
Furthermore, by combining the relations in (\ref{SS1}) with those 
in (\ref{S9}) and (\ref{SS2}), we find 
\be
\label{S11}
n_s - 1 \sim -3 \epsilon\, , \quad 
\alpha_s \sim -3 \left( \frac{H'(N)}{H(N)} \right)^2 
\sim -3 \epsilon^2 \, .
\ee
We should note that the running of the spectral index $\alpha_\mathrm{s}$ is not always small and negative.

\section{Scalar field theories compatible with the Planck and BICEP2 results}

We study a scalar field theory consistent with the BICEP2 result and 
that compatible with the Planck data in terms of the tensor-to-scalar 
ratio $r$ by taking account of the running of the spectral index 
$\alpha_\mathrm{s}$. 

According to the Planck analysis~\cite{Ade:2013lta, Ade:2013uln}, 
$n_{\mathrm{s}} = 0.9603 \pm 0.0073\, (68\%\,\mathrm{CL})$ 
and $\alpha_\mathrm{s} = -0.0134 \pm 0.0090\, (68\%\,\mathrm{CL})$ 
with the Planck and WMAP~\cite{WMAP, Hinshaw:2012aka} data. The sign of $\alpha_\mathrm{s}$ is negative at 
$1.5 \sigma$ level. Furthermore, $r< 0.11\, (95\%\,\mathrm{CL})$. 
On the other hand, the result of the BICEP2 experiment is 
$r = 0.20_{-0.05}^{+0.07}\, (68\%\,\mathrm{CL})$~\cite{Ade:2014xna} 
(for very recent discussions on the data analysis of the foreground data 
related to the BICEP2 result, see, e.g., Refs.~\cite{A-A, Mortonson:2014bja}). 

We here remark that 
from the consequences of the reconstruction in (\ref{S11}) with Eq.~(\ref{S11}), e.g., for $\epsilon = 1.1 \times 10^{-2}$, we acquire 
$n_{\mathrm{s}} = 0.967$, $r=0.176$, and $\alpha_\mathrm{s} = -3.63 \times 10^{-4}$. Hence, it is interpreted that the reconstruction procedure could 
lead to the scalar field theories realizing inflation compatible with the Planck and BICEP2 results. 

In inflationary models with a single inflaton filed $\phi$, 
the Lyth bound~\cite{Lyth:1996im} has been known regarding the difference of the inflaton amplitude $\phi_\mathrm{i}$ at the initial time $t_\mathrm{i}$ of 
inflation from that $\phi_\mathrm{f}$ at the final time $t_\mathrm{f}$ of it. 
It can be represented as $\Delta \left(\kappa \phi \right) \equiv \left|\kappa\phi_\mathrm{i} - \kappa\phi_\mathrm{f} \right| \gtrsim N \sqrt{r/8}$~\cite{Hossain:2014ova}. For $N = 50 \, (60)$ and $r = 0.10$, we have $\Delta \left(\kappa \phi \right) \gtrsim 5.6 \, (6.1)$~\cite{Antusch:2014cpa}. When $r \geq [<] \, 3.2 \, (2.0) \times 10^{-3}$ with $N= 50 \, (60)$, 
we find $\Delta \left(\kappa \phi \right) \geq [<] \, 1$. 
Consequently, in what follows, as the condition for an inflation model consistent with the BICEP2 result, we consider $\Delta \left(\kappa \phi \right) \geq 1$, whereas as that compatible with the Planck data, we regard $\Delta \left(\kappa \phi \right) < 1$.

\subsection{Inflationary model consistent with the BICEP2 result}

First, we explore the so-called chaotic inflation model~\cite{Linde:1981mu}, where the inflaton potential of a canonical scalar field $\phi$ is given by 
\begin{equation} 
V(\phi) = \bar{V}_\mathrm{c} \left( \kappa \phi \right)^p \,, 
\label{eq:3.1}
\end{equation}
with $\bar{V}_\mathrm{c}$ and $p$ constants. In this model, from equations in (\ref{S2}) 
we obtain
\begin{equation} 
\epsilon \simeq \frac{p}{4N+p}\,, \quad 
\eta \simeq \frac{2\left(p-1\right)}{4N+p}\,, \quad 
\xi^2 \simeq \frac{4\left(p-1\right)\left(p-2\right)}{\left(4N+p\right)^2}\,. 
\label{eq:3.2}
\end{equation}
In deriving these relation, we have used $\kappa \phi \simeq \sqrt{p\left(4N+p\right)/2}$. This comes from the condition of 
$\epsilon (\phi_\mathrm{f}) < 1$, 
leading to $\phi > p/\left(\sqrt{2} \kappa\right)$, 
and the relation 
$N= -\int_{t_\mathrm{f}}^t H(\hat{t}) d \hat{t} \approx 
\kappa^2 \int_{\phi_\mathrm{f}}^{\phi} \left( V(\hat{\phi})/V'(\hat{\phi}) \right) d \hat{\phi}$, where the approximate equality follows from the relation 
$H \simeq -\kappa \left( V(\phi)/V'(\phi) \right) \left( \kappa \dot{\phi} \right)$ under the slow-roll approximation. 

With Eq.~(\ref{S3}) and the equations in (\ref{SS1}), 
we acquire
\begin{equation} 
n_\mathrm{s} \simeq \frac{4\left(N-1\right)-p}{4N+p}\,, \quad 
r \simeq \frac{16p}{4N+p} \,, \quad 
\alpha_\mathrm{s} \simeq -\frac{8\left(p+2 \right)}{\left(4N+p\right)^2}\,.
\label{eq:3.3}
\end{equation}
For instance, when $(N, p)=(50, 2)$ and $(60, 3)$, we get 
$(n_\mathrm{s}, r, \alpha_\mathrm{s}) = (0.960, 0.158, -7.84 \times 10^{-4})$ 
and $(0.959, 0.198, -6.77 \times 10^{-4})$, respectively. 
Accordingly, this model could be considered to be consistent with the BICEP result, even though 
the absolute value of $\alpha_\mathrm{s}$ is smaller than that of the Planck one.

\subsection{Inflationary model compatible with the Planck analysis}

Next, we investigate the so-called natural (or axion) inflation model~\cite{Freese:1990rb} with the following inflaton potential 
of a pseudo-scalar field $\phi$ such as a pseudo Nambu-Goldstone boson
\begin{equation} 
V(\phi) = \bar{V}_\mathrm{a} \left[ 1+ \cos \left(\frac{\phi}{f_\mathrm{a}} \right) \right]\,,
\label{eq:3.4}
\end{equation}
where $\bar{V}_\mathrm{a}$ and 
$f_\mathrm{a}$ are constants, and 
$\phi/f_\mathrm{a}$ is a dimensionless combination. 
The number of $e$-folds could be represented as~\cite{Gong:2014cqa}
\begin{equation} 
N = 2\kappa^2 f_\mathrm{a}^2 
\ln \left[ \frac{\sin \left( \phi_\mathrm{f}/f_\mathrm{a} \right)}{\sin \left( \phi_\mathrm{h}/f_\mathrm{a} \right) } \right]\,,
\label{eq:3.8}
\end{equation}
where $\phi_\mathrm{h}$ is the amplitude of $\phi$ at the time 
$t_\mathrm{h}$ when the curvature perturbation with the wave number $k$ first crosses the horizon during inflation. 
In this case, 
$N$ is considered to be the value of the number of $e$-folds for the 
the curvature perturbation with the wave number $k$. 
Moreover, from a relation $\epsilon (\phi_\mathrm{f})=1$ at the end of 
inflation, we see that 
$\phi_\mathrm{f}/f_\mathrm{a}$ is described as~\cite{Gong:2014cqa} 
\begin{equation} 
\frac{\phi_\mathrm{f}}{f_\mathrm{a}} 
= \arccos \left[ \frac{1-2\left( \kappa f_\mathrm{a} \right)^2}{1+2\left( \kappa f_\mathrm{a} \right)^2} \right]\,. 
\label{eq:3.9}
\end{equation}
For this model, it follows from equations in (\ref{S2}) that
\begin{eqnarray}
\epsilon \Eqn{=} \frac{1}{2\kappa^2 f_\mathrm{a}^2} \frac{\sin^2\left(\phi_\mathrm{h}/f_\mathrm{a}\right)}{\left[1+\cos\left(\phi_\mathrm{h}/f_\mathrm{a}\right)\right]^2} \,, 
\label{eq:3.5} \\
\eta \Eqn{=} - \frac{1}{\kappa^2 f_\mathrm{a}^2} \frac{\cos\left(\phi_\mathrm{h}/f_\mathrm{a}\right)}{1+\cos\left(\phi_\mathrm{h}/f_\mathrm{a}\right)} \,,
\label{eq:3.6} \\
\xi^2 \Eqn{=} - \frac{1}{\kappa^4 f_\mathrm{a}^4} \frac{\sin^2\left(\phi_\mathrm{h}/f_\mathrm{a}\right)}{\left[1+\cos\left(\phi_\mathrm{h}/f_\mathrm{a}\right)\right]^2} \,, 
\label{eq:3.7} 
\end{eqnarray}
where $\phi_\mathrm{h}$ is described by Eq.~(\ref{eq:3.8}) with $N$ and 
$\phi_\mathrm{f}$ as $\sin \left( \phi_\mathrm{h}/f_\mathrm{a} \right)
= \sin \left( \phi_\mathrm{f}/f_\mathrm{a} \right) \exp \left[-N/\left( 2\kappa^2 f_\mathrm{a}^2\right) \right]$. 
By using this equation, Eq.~(\ref{S3}), and equations in (\ref{SS1}), we find 
\begin{eqnarray} 
n_\mathrm{s} \Eqn{=} 
1-\frac{1}{\kappa^2 f_\mathrm{a}^2} 
\frac{1}{\left[1+\cos\left(\phi_\mathrm{h}/f_\mathrm{a}\right)\right]^2} 
\left\{ \sin^2\left(\frac{\phi_\mathrm{h}}{f_\mathrm{a}}\right) 
+2 \left[ 1+ \cos\left(\frac{\phi_\mathrm{h}}{f_\mathrm{a}}\right) \right] 
\right\}\,, 
\label{eq:3.10} \\
r \Eqn{=}  
\frac{8}{\kappa^2 f_\mathrm{a}^2} \frac{\sin^2\left(\phi_\mathrm{h}/f_\mathrm{a}\right)}{\left[1+\cos\left(\phi_\mathrm{h}/f_\mathrm{a}\right)\right]^2} \,, 
\label{eq:3.11} \\
\alpha_\mathrm{s} \Eqn{=} 
-\frac{4}{\kappa^4 f_\mathrm{a}^4} 
\frac{\sin^2\left(\phi_\mathrm{h}/f_\mathrm{a}\right)}{\left[1+\cos\left(\phi_\mathrm{h}/f_\mathrm{a}\right)\right]^3} \,.
\label{eq:3.12}
\end{eqnarray}
The amplitude of $\phi$ at the first horizon crossing time 
for the number of $e$-folds $N=50$--$60$ before the 
end of inflation is observed in the fluctuations 
in terms of the cosmic microwave background (CMB) radiation. 
As an example, for 
$(N, \kappa f_\mathrm{a})=(50.0, 5.00)$ and $(60.0, 5.50)$, 
we have $(n_\mathrm{s}, r, \alpha_\mathrm{s}) = (0.960, 8.37 \times 10^{-4}, -8.39 \times 10^{-6})$ and $(0.967, 5.85 \times 10^{-4}, -4.84 \times 10^{-6})$, respectively. 
In evaluating these numerical values, we have also used 
the related equations 
$\cos \left( \phi_\mathrm{h}/f_\mathrm{a} \right) = \sqrt{1- \sin^2 \left( \phi_\mathrm{f}/f_\mathrm{a} \right) \exp \left[-N/\left( \kappa^2 f_\mathrm{a}^2\right) \right]}$ and $\sin \left( \phi_\mathrm{h}/f_\mathrm{a} \right) = 
2\sqrt{2} \kappa f_\mathrm{a}/\left[1+ 2\left( \kappa f_\mathrm{a} \right)^2 \right] $ following from Eq.~(\ref{eq:3.9}). 
Consequently, this model can be regarded to be compatible with the Planck result.

\subsection{Model with a hyperbolic inflaton potential}

Furthermore, as a model of inflation, 
we explore the action of a scalar field theory 
in (\ref{S1}) with the following potential of the inflaton $\phi$~\cite{Yokoyama:1987an}: 
\begin{equation} 
V(\phi) = \bar{V}_\mathrm{n} \cosh \left( \gamma_\mathrm{n} 
\kappa \phi \right) \,, 
\label{eq:3C.1}
\end{equation}
where $\bar{V}_\mathrm{n}$ and $\gamma_\mathrm{n}$ are constants. 
In this model, the number of $e$-folds during inflation reads 
\begin{equation}  
N = - \int_{t_\mathrm{f}}^t H(\hat{t}) d \hat{t} \simeq \kappa^2 \int_{\phi_\mathrm{f}}^{\phi} \frac{V(\hat{\phi})}{V'(\hat{\phi})} d \hat{\phi} 
\approx \ln \left[ \frac{\left( \gamma_\mathrm{n} \kappa \phi \right)}{\left( \gamma_\mathrm{n} \kappa \phi_\mathrm{f} \right)} \right]\,, 
\label{eq:3C.2}
\end{equation}
where in deriving the last approximate equality, 
we have used $\gamma_\mathrm{n} \kappa \phi \ll 1$ with 
$\gamma_\mathrm{n} \ll 1$. 
In what follows, we concentrate on this regime. 
We remark that for $\gamma_\mathrm{n} \kappa \phi \ll 1$, 
the inflaton potential in Eq.~(\ref{eq:3C.1}) becomes 
$V(\phi) \approx \bar{V}_\mathrm{n} \left\{1 + \left(1/2\right) \left( \gamma_\mathrm{n} \kappa \phi \right)^2 \right\}$. Such a form is similar to that of the so-called hybrid inflation model~\cite{Linde:1993cn} at the inflationary stage. Here, we only examine the inflationary stage and consider that the graceful exit from inflation occurs due to some additional mechanism. 
It follows from the relations in (\ref{S2}) and Eq.~(\ref{eq:3C.2}) that 
\begin{eqnarray}
\epsilon \Eqn{=} \frac{\gamma_\mathrm{n}^2}{2} 
\tanh^2 \left( \gamma_\mathrm{n} \kappa \phi \right)
\approx \frac{\gamma_\mathrm{n}^2}{2} 
\tanh^2 \left[ \gamma_\mathrm{n} \kappa \phi_\mathrm{f} \exp \left( N \right) \right]\,, 
\label{eq:3C.3} \\
\eta \Eqn{=} \gamma_\mathrm{n}^2\,,
\label{eq:3C.4} \\
\xi^2 \Eqn{=} \gamma_\mathrm{n}^4 \tanh^2 
= 2 \gamma_\mathrm{n}^2 \epsilon 
\approx \gamma_\mathrm{n}^4 
\tanh^2 \left[ \gamma_\mathrm{n} \kappa \phi_\mathrm{f} \exp \left( N \right) \right]\,. 
\label{eq:3C.5}
\end{eqnarray}
{}From Eq.~(\ref{S3}), the equations in (\ref{SS1}), and 
Eqs.~(\ref{eq:3C.3})--(\ref{eq:3C.5}), we obtain 
\begin{eqnarray} 
n_\mathrm{s} \Eqn{\sim}
1+2\gamma_\mathrm{n}^2 -3\gamma_\mathrm{n}^2 
\tanh^2 \left( \gamma_\mathrm{n} \kappa \phi \right)
\approx 1+2\gamma_\mathrm{n}^2 -3\gamma_\mathrm{n}^2 
\tanh^2 \left[ \gamma_\mathrm{n} \kappa \phi_\mathrm{f} \exp \left( N \right) \right]\,, 
\label{eq:3C.6} \\
r \Eqn{\sim} 8 \gamma_\mathrm{n}^2 \tanh^2 \left( \gamma_\mathrm{n} \kappa \phi \right)
\approx 8 \gamma_\mathrm{n}^2 \tanh^2 \left[ \gamma_\mathrm{n} \kappa \phi_\mathrm{f} \exp \left( N \right) \right]\,, 
\label{eq:3C.7} \\
\alpha_\mathrm{s} \Eqn{\sim} 6 \gamma_\mathrm{n}^4 \frac{\tanh^2 \left( \gamma_\mathrm{n} \kappa \phi \right)}{\cosh^2 \left( \gamma_\mathrm{n} \kappa \phi \right)} 
\approx 
6 \gamma_\mathrm{n}^4 \frac{\tanh^2 \left[ \gamma_\mathrm{n} \kappa \phi_\mathrm{f} \exp \left( N \right) \right]}{\cosh^2 \left[ \gamma_\mathrm{n} \kappa \phi_\mathrm{f} \exp \left( N \right) \right]}\,.
\label{eq:3C.8}
\end{eqnarray}
For example, 
when $(N, \gamma_\mathrm{n}, \kappa \phi_\mathrm{f})=(50, 3.00 \times 10^{-2}, 
1.04 \times 10^{-20})$ 
and $(60, 4.00 \times 10^{-2}, 3.01 \times 10^{-25})$, 
we find $(n_\mathrm{s}, r, \alpha_\mathrm{s}) = 
(0.999, 6.17 \times 10^{-3}, 5.98 \times 10^{-7})$ and $(0.999, 9.90 \times 10^{-3}, 2.69 \times 10^{-6})$, respectively. 
For the latter case, 
the value of $r$ becomes about $\mathcal{O}(0.01)$, although it is 
smaller than the BICEP by one order. 

We finally mention that 
when equations in (\ref{S7}) can be solved with respect to $H$, we can 
explicitly get the corresponding scalar field theory. 
Indeed, however, this is too complicated calculation to be executed. 
That is why the investigations are limited here by presenting several 
simple examples. 

\section{Summary}

In summary, we have reconstructed the scalar field models to explain the BICEP2 and Planck results by taking into consideration the running of the spectral index. As concrete examples, we have explored 
the chaotic inflation, natural (or axion) inflation, 
and a model of inflation in which the inflaton potential 
has a hyperbolic form. 

Our method demonstrated in this Letter is considered to be 
an effective way of making scalar field models to 
account for any observational data taken from not only the current BICEP2 and Planck but also the future observational missions. 
Through this process, in principle, wider classes of the inflaton potentials which can realize inflation compatible with observations could be built. 
Consequently, it is expected that this procedure would lead to novel insights on the inflaton potentials to yield the consequences explaining any observations including the BICEP2 experiment and the Planck satellite. 
Furthermore, recently there have been raised some doubts on the the prediction 
of the BICEP2 experiment. Even if this is true, our method gives 
the possibility to reconstruct scalar field theories realizing 
viable inflation from the known values of slow-roll inflationary parameters. 
With release of more observational results, this reconstruction method may give more exact theoretical models of inflation. 

In addition, very recently, an attempt of building $F(R)$ gravity theories~\cite{R-NO-CF-CD} with inflation from observational data has been executed in Refs.~\cite{Codello:2014sua, SCMOZ-RCVZ} (see also Ref.~\cite{BMOS-BCOZ}). As our next step, the formulations of reconstruction procedures of modified gravity theories such as $F(R)$ gravity consistent with the BICEP2 and Planck data will be developed in the future continuous work.

\section*{Acknowledgments}

The work has been supported in part by Russian Ministry of Education and Science, project TSPU-139 (S.D.O.), the JSPS Grant-in-Aid 
for Scientific Research (S) \# 22224003 and (C) \# 23540296 (S.N.), 
and that for Young Scientists (B) \# 25800136 (K.B.).


\end{document}